\documentclass[aps,prd,reprint,superscriptaddress,nofootinbib,longbibliography]{revtex4-2}

\pdfoutput=1

\usepackage[T1]{fontenc}
\usepackage[utf8]{inputenc}

\usepackage{amsmath,amssymb,amsthm,mathtools}
\usepackage{mathrsfs}
\usepackage{bm}

\usepackage{graphicx}
\usepackage{subfig}

\usepackage[
colorlinks=true,
linkcolor=blue,
citecolor=blue,
urlcolor=blue
]{hyperref}

\begin{document}

\title{Asymptotic regularization method: A constructive approach}

\author{Christian Dur\'an Romero}
\email{chduran@ucm.es}

\author{Luis J. Garay}
\email{luisj.garay@ucm.es}

\author{Mercedes Mart\'in-Benito}
\email{m.martin.benito@ucm.es}

\affiliation{Departamento de F\'isica Te\'orica and IPARCOS, Universidad Complutense de Madrid, Plaza de las Ciencias 1, 28040 Madrid, Spain
}

\author{Rita B. Neves}
\email{rita.neves@sheffield.ac.uk}

\affiliation{School of Mathematical and Physical Sciences, University of Sheffield, Hicks Building, Hounsfield Road, Sheffield S3 7RH, United Kingdom}

\begin{abstract}
We introduce a new regularization scheme for divergent integrals in quantum field theory.
The framework is based on the structural decomposition of the integrand asymptotic expansion, which distinguishes between contributions that drive ultraviolet singularities and those that remain finite.
This asymptotic regularization method isolates the genuinely singular sector and enables a consistent subtraction of divergences while maintaining covariance and gauge symmetry. In single-scale theories, we show that the renormalized quantities exhibit a non-local logarithmic dependence uniquely determined by the ultraviolet asymptotics, offering a derivation of logarithmic terms that is independent of standard renormalization-group flows. Because it relies only on asymptotic structure rather than on standard relativistic power counting, the method is naturally applicable to theories with modified dispersion relations and non-standard ultraviolet scaling.
Although formulated here for ultraviolet divergences, the underlying strategy extends straightforwardly to infrared singularities.
\end{abstract}

\maketitle

\section{Introduction}
\label{sec:introduction}

Ultraviolet (UV) divergences are a structural feature of quantum field theory (QFT) and must be regularized in order to define renormalized amplitudes and extract physically meaningful quantities \cite{Collins:105730}. A variety of regularization schemes have been developed for this purpose, among which dimensional regularization has become the standard framework. Its algebraic simplicity, compatibility with gauge invariance and Lorentz symmetry, and its natural connection with renormalization group techniques make it particularly well suited for perturbative QFT. In this scheme, the spacetime dimension is analytically continued and UV singularities appear as poles in the dimension parameter \cite{Bollini:1972ui, tHooft:1972tcz}.

UV divergences are not exclusive to  QFT in flat spacetime. They also arise in QFT in curved spacetimes, where the absence of global symmetries and the presence of background curvature introduce additional subtleties \cite{BirrellDavies1982}. Although dimensional regularization can often be applied in covariant formulations, its implementation may become technically involved or conceptually less transparent when the UV structure of the theory departs from the standard relativistic scaling \cite{LopezNacir:2009bhs}.

A particularly relevant example occurs in the presence of modified dispersion relations (MDRs). Such modifications arise in a variety of contexts, including effective field theories, Lorentz-violating extensions, and quantum-gravity-inspired models. By construction, MDRs alter the high-momentum behavior of propagators and therefore modify the UV asymptotic structure of integrands~\cite{Unruh1976db, corley1996}. Since dimensional regularization is primarily designed to treat divergences within the framework of standard relativistic power counting, changes in the asymptotic scaling may render its direct application insufficient to fully capture or isolate the UV structure in a transparent manner. When the UV behavior deviates from standard relativistic power counting, the relation between asymptotic scaling, pole structure, and renormalization-group interpretation becomes less direct. This suggests that the regularization procedure should be formulated directly in terms of the UV behavior. What is needed is a scheme whose applicability does not rely on standard relativistic scaling, but only on the asymptotic structure itself.

These considerations motivate the search for regularization schemes that make the UV asymptotic behavior of the integrand explicit. Several established approaches share this philosophy. For instance, the BPHZ method implements subtractions determined by local Taylor expansions of amplitudes \cite{Bogoliubov:1957gp}. Adiabatic subtraction in cosmological spacetimes isolates UV contributions through asymptotic expansions of mode functions \cite{parker_toms_qft_curved_spacetime}. Another example is the method of regions, that systematically separates contributions associated with different scaling regimes \cite{Beneke:1997zp}. All these techniques exploit, in one way or another, the fact that UV divergences are encoded in the large-momentum structure of the integrals.

Nevertheless, although these approaches make essential use of the UV asymptotic structure, they do not always provide a formulation that isolates the singular behavior in terms of an explicit analytic regulator. In particular, subtraction-based schemes such as BPHZ or adiabatic renormalization remove divergent contributions directly at the integrand level but do not naturally lead to a structure characterized by poles in a regulator parameter, as occurs in dimensional regularization. Likewise, methods such as the method of regions typically yield effective or asymptotic expansions adapted to specific scaling limits, rather than exact analytic representations of the full amplitude. As a consequence, although these approaches can be implemented in a manner consistent with Lorentz invariance (and, when treated appropriately, with gauge symmetry), the connection with renormalization-group concepts or with a universal pole structure may be less transparent. 

The purpose of this work is to formulate and analyze a regularization scheme that makes the role of the asymptotic structure explicit while preserving the infrared (IR) structure as well as symmetries of the non regularized theory. The method that we propose exploits the large-momentum expansion of the integrand isolating the divergent part. In this manner, the scheme applies to a broad class of integrands determined by their asymptotic behavior, yet maintains the structural features, such as analytic control and compatibility with renormalization group. In standard relativistic settings, our construction reproduces the usual well-known singular structure. However, because it is formulated directly in terms of asymptotic behavior, it remains applicable even when the UV scaling departs from the relativistic one.

Our method, which we refer to as \emph{asymptotic regularization}, rests on a simple but structural observation: the UV behavior of a rotationally invariant integral is encoded in the large-momentum radial expansion of its integrand, yet the different terms in this asymptotic series do not play equivalent roles. In particular, only specific components of the asymptotic expansion are responsible for genuine UV singular structures, while others contribute in a qualitatively different manner. Rather than regularizing the full integral in a uniform way, we exploit this hierarchy. The method separates the integral into distinct contributions according to their asymptotic behavior: a finite part containing the IR-sensitive information, a set of terms with different UV scaling, and a marginal component responsible for the logarithmic divergence. Each sector is then treated according to its structural role. In this way, the UV singular structure is isolated in a transparent manner, while the remaining contributions are handled directly at the physical dimension. While the present work focuses on UV divergences, the same construction applies to IR singularities by analyzing the small-momentum behavior of the integrand.

Furthermore, under the assumptions described above, the analysis reveals a remarkable constrained behavior: when a single physical scale is present, the regularized quantity necessarily develops a non-local logarithmic dependence on this scale whose coefficient is entirely determined by the UV asymptotics. This provides a direct UV origin of non-local logarithmic terms, independently of renormalization-group arguments. An analogous statement holds for IR singularities, where the same marginal structure controls the logarithmic behavior, reinforcing the universality of the asymptotic framework.

When no explicit hard cutoff is introduced to regularize what we identify with the only singular structure of the divergent integrals, gauge identities such as Ward and Slavnov–Taylor identities \cite{Ward:1950xp, Takahashi:1957xn, Taylor:1971ff, Slavnov:1972fg} are preserved provided the procedure is implemented consistently. Lorentz invariance remains manifest throughout the construction, and the framework is compatible with covariant formulations, including those relevant for quantum field theory in curved spacetimes. At the same time, since our method targets only the UV structures and does not rely on full spacetime covariance, the method can be consistently extended to more general settings in which covariance is not manifest, such as theories with MDRs, providing a unified framework for treating UV divergences both within and beyond the standard relativistic paradigm.

This paper is organized as follows. In Sec.~\ref{Section II} we review the analytic regularization of algebraic singularities and establish the basic properties of power-law divergences. In Sec.~\ref{Section III.A} we formulate the asymptotic regularization prescription, computing the isolated singular structure with different regulators in Sec.~\ref{Section III.B}. UV control of non-local logarithmic terms in presence of a scale is studied in Sec.~\ref{Section III.C}. In Sec.~\ref{Section III.D} we present an alternative derivation of these non-local logarithms. In Sec.~\ref{Section III.E} we present explicit examples that illustrate the implementation of the method and clarify its relation to standard dimensional regularization. A detailed comparison between asymptotic regularization and other existing approaches is presented in Sec.~\ref{Section IV}. In Sec.~\ref{Section IV.A} we first compare asymptotic regularization with dimensional regularization from the viewpoint of symmetry preservation. In Sec.~\ref{Section IV.B} we then show how the pole can be isolated and studied directly within the asymptotic regularization framework. Then, in Sec.~\ref{Section IV.C} we analyze an example that lies beyond the direct scope of other regularization schemes, illustrating how the method can be applied in situations where standard techniques are not readily available.  Finally, Sec.~\ref{Section V} contains our conclusions. IR singularities are discussed separately in Appendix~\ref{app A}, where the extension of the method is presented for completeness.  Appendix~\ref{app B} is devoted to a study of a class of regulators that can be employed to the isolated divergent part.

\section{Regularization of UV divergences}
\label{Section II}

In this section we show the main mathematical results that set the basis for the regularization method that will be presented in the next section.

\subsection{Setup and hypotheses}
\label{Section IIx}

We consider integrals over a $D$-dimensional Euclidean momentum space,
\begin{equation}
 \int_{\mathbb{R}^D} d^D\ell\, f(\ell).
\label{eq:I_def}
\end{equation}
where $\ell = |\vec{\ell}|$ denotes the Euclidean norm and the scalar function $f$ is assumed to be rotationally invariant, so that it only depends on $\ell$. This assumption entails no loss of generality: by standard arguments, any tensorial or non-spherically symmetric integrand can be reduced to a finite linear combination of scalar radial integrals after angular averaging~\cite{Collins:105730}. In particular, integrals involving products of components of $\vec{\ell}$ reduce to scalar integrals multiplied by combinations of the Euclidean metric, by rotational invariance.

Throughout this work we restrict attention to integrals exhibiting UV divergences. The regularization procedure will therefore be formulated so as to control the singular behavior responsible for such divergences. If IR divergences were present, they could be treated by applying the same strategy after introducing a suitable IR regulator. The resulting regularized structure would again be determined by the algebraic nature of the underlying singularity. 

 After angular integration, any $D$-dimensional integral of a rotationally invariant function reduces to a one-dimensional radial integral. This procedure is formally grounded in the set of axioms for $D$-dimensional integration (namely linearity, translation invariance, and scaling) as defined in Ref.~\cite{Collins:105730}. Under these prescriptions, the integral \eqref{eq:I_def} takes the form
\begin{equation}
 \Omega_D \int_0^\infty d\ell\, \ell^{D-1} f(\ell),
\qquad
\Omega_D \equiv \frac{2\pi^{D/2}}{\Gamma(D/2)},
\label{eq:I_radial_final}
\end{equation}
where $\Omega_D$ denotes the volume of the unit $D-1$ sphere. Notice that the UV behavior of the original $D$-dimensional integral is entirely controlled by the large-$\ell$ asymptotics of the associated radial integrand.  Although the explicit form of $\Omega_D$ plays no role at this stage for purely scalar integrals, we retain it for later consistency. 
In gauge theories, dimensional continuation must be implemented coherently at the level of angular integrations, and keeping $\Omega_D$ ensures that tensorial structures and their associated $D$-dependence are treated in a way compatible with gauge identities \cite{Collins:105730}.

\subsection{Analytic regularization and algebraic singularities}The theory of generalized functions with algebraic singularities provides a systematic
framework to assign a precise meaning to otherwise divergent expressions by embedding
them into families depending on a complex parameter.  By algebraic singularity we mean that, as $\ell$ approaches a singular point $\ell_0$, the function is bounded by a finite power of $|\ell-\ell_0|^{-1}$. In other words, its growth is at most polynomial \cite{GelfandShilov1964GF}.

Within this approach, quantities that are ill-defined at specific values of a parameter
are first defined in a suitable domain of convergence and subsequently extended to the
complex plane by analytic (or, more generally, meromorphic) continuation. The resulting
meromorphic structure encodes the divergences of the original expression. In particular, distributions depending analytically on a complex exponent
admit a unique meromorphic continuation, and non-integrable power-law behaviors give rise
to simple poles at specific values of the exponent.
This general mechanism underlies many regularization schemes commonly used in mathematical
physics, including analytic and zeta-function regularization~\cite{GelfandShilov1964GF}.

We will restrict attention to integrands whose analytic continuation exhibits a controlled singular behavior. For these functions, divergences appear as isolated poles whose position and order are determined by the algebraic components of the asymptotic expansion. In this restricted sense, the singular behaviour is universal within the specified algebraic case, and does not depend on the detailed subleading structure of the integrand.

We now illustrate these general ideas with a simple explicit example, which, while fully consistent with these general results, is worked out explicitly here for later use.

\subsubsection*{Example: A basic algebraic divergence}
Let us consider the  following divergent integral:
\begin{equation}
   \mathcal I(a)=\int_0^\infty d\ell\, \ell^a,
   \qquad a\geq -1.
   \label{eq:I_integral}
\end{equation} 
In order to analyze its singular structure in a systematic way, we   complexify the exponent and   study  
  the behavior, as $\lambda \to 0 ^{+}$, of the integral
 \begin{equation}
    \mathcal J(s;\lambda)\equiv\int_\lambda^\infty d\ell\,\ell^s,
    \qquad s\in\mathbb{C}, \lambda \in \mathbb{R}^{+}.
    \label{eq:J_integral}
\end{equation}

This integral converges absolutely for $\Re(s)<-1$, while it diverges for $\Re(s)\geq -1$. The usual strategy is to use the convergence region to define the integral elsewhere by analytic continuation.
To make this explicit, we split the integration domain at an arbitrary scale $\lambda_*>\lambda$,
\begin{equation}
    \mathcal J(s;\lambda)
    =\int_\lambda^{\lambda_*} d\ell\,\ell^s
    +\int_{\lambda_*}^\infty d\ell\,\ell^s.
    \label{eq:J_split}
\end{equation}

The first term is finite and can be evaluated directly,
\begin{equation}
    \int_\lambda^{\lambda_*} d\ell\,\ell^s
    =\frac{\lambda_*^{s+1}}{s+1}-\frac{\lambda^{s+1}}{s+1} .
    \label{eq:J_split_finite}
\end{equation}

The second term, however, diverges when $\Re(s)\geq -1$. Nevertheless, for $\Re(s)<-1$ it converges absolutely and can be computed explicitly,
\begin{equation}
   \int_{\lambda_*}^\infty d\ell\,\ell^s
   =-\frac{\lambda_*^{s+1}}{s+1},\qquad \Re(s)<-1.
   \label{eq:J_split_tail}
\end{equation}
The right-hand side of the previous result defines a meromorphic function of $s$ on the entire complex plane, with a single simple pole at $s=-1$.
We now use this expression to define the divergent contribution in the region $\Re(s)\geq -1$ by analytic continuation. 

The analytic continuation yields
\begin{equation}
   \mathcal J(s;\lambda)
   =-\frac{\lambda^{s+1}}{s+1},
   \qquad s\in\mathbb{C}\setminus\{-1\}.
   \label{eq:J_meromorphic}
\end{equation}
By the uniqueness theorem for analytic continuation, this expression provides the unique analytical extension of $\mathcal J(s;\lambda)$ from the convergence domain $\Re(s)<-1$ to the whole complex plane except  $s=-1$. In particular, no additional singularities can appear in the continuation.
It follows immediately from Eq.~\eqref{eq:J_meromorphic} that $\mathcal J(s;\lambda)$ possesses a single simple pole at $s=-1$. Indeed, since the numerator $\lambda^{s+1}$ is an entire function and satisfies $\lambda^{0}=1\neq 0$, no cancellation occurs at $s=-1$, and the pole is therefore genuine.

Therefore, for all $s> -1$ the analytic continuation satisfies
\begin{equation}
   \lim_{\lambda\to 0^+}\mathcal J(s;\lambda)=0.
\end{equation}
Thus, for $\Re(s)\geq-1$, the unique value of the exponent for which a non-vanishing singular contribution survives in the limit $\lambda\to 0^+$ is $s=-1$.

The analysis above provides a natural regularization prescription. 
We define the regularized value of $\mathcal I(a)$ by complexifying the exponent, computing the meromorphic continuation of $\mathcal J(s;\lambda)$, and then evaluating the result at $s=a$. More precisely, we set
\begin{equation}
   \mathcal I_{\mathrm{reg}}(a)
   \equiv \lim_{\lambda\to 0^+}\,\mathcal J(s;\lambda)\big|_{s=a}, \qquad a\geq-1,
\end{equation}
where $\mathcal J(s;\lambda)$ is understood as its analytic continuation given in \eqref{eq:J_meromorphic}.

For $a  > -1$ the regularized value vanishes, i.e.,
\begin{equation}
  \mathcal I_{\mathrm{reg}}(a)=0, \qquad a > -1.
\end{equation}

At $a=-1$, however, the meromorphic continuation exhibits a simple pole,
\begin{equation}
   \mathcal I_{\mathrm{reg}}(a)
   \sim -\frac{1}{a+1}
   \qquad \text{as } a\to -1,
\end{equation}
since,  for $a  \to -1$, $\lambda^{a+1}\to 1$. 
The divergence of the original integral is therefore encoded in the simple pole at $a=-1$.

We conclude that analytic continuation provides a consistent regularization of the divergent integral \eqref{eq:I_integral}: all power-law divergences are set to zero, while logarithmic divergences are encoded as simple poles of the analytically continued expression.

This example provides a concrete realization of the general results on algebraic
singularities and analytic regularization discussed in the theory of generalized
functions \cite{GelfandShilov1964GF}.

\section{Asymptotic regularization}
\label{Section III}

In this section we introduce the regularization scheme that will be employed throughout this work, which we call \textit{asymptotic regularization}. 
The central idea of the method is that the UV behavior of a divergent integral can be systematically organized according to the large-momentum asymptotic structure of its integrand. 
Different asymptotic sectors play qualitatively distinct roles in the physics of the problem: Some contributions encode finite or IR-sensitive information; others generate UV growth that can be treated unambiguously; and a marginal sector which, when present, is solely responsible for the genuine logarithmic divergence.

\subsection{The method}
\label{Section III.A}

Let us consider a rotationally invariant, $D$-dimensional integral of the form~\eqref{eq:I_radial_final}. In what follows, we focus on the analysis of UV-divergent integrals. IR singularities, when present, can be handled in a completely analogous manner within the same asymptotic framework, as outlined in Appendix~\ref{app A}. 
Independently of the detailed functional form of $f$, its large-$\ell$ behavior allows for a decomposition into asymptotically distinct contributions. 
Schematically, for sufficiently large $\ell$, we may write
\begin{equation}
    f(\ell) \sim f_\text{div}(\ell)+ R(\ell), \quad f_\text{div}(\ell)\equiv f_{\mathrm{UV}}(\ell)
    + f_{\mathrm{marg}}(\ell),
    \label{eq: f decomposition}
\end{equation}
where $f_\text{div}$ captures all divergent contributions. Here, $f_{\mathrm{UV}}$ collects the non-marginal asymptotic contributions. If present, the marginal term $f_{\mathrm{marg}}$ takes the form
\begin{equation}
    f_\text{marg}(\ell)= b\,\ell^{-D}, \qquad b\in \mathbb{R}.
    \label{eq: f marg}
\end{equation}
Finally, the remainder $R(\ell)$ decays strictly faster than $\ell^{-D}$ in the UV.

This separation is purely asymptotic, yet it has direct consequences for the singular structure of the integral. 
Contributions whose asymptotic scaling differs from $\ell^{-D}$ do not generate logarithmic UV singularities. 
They may lead to finite contributions or to divergences of a different nature, depending on their precise structure. The logarithmic divergence is uniquely controlled by the marginal term.
As we shall see, this contribution gives rise to a genuine simple pole in the regularized expression that cannot be eliminated by purely algebraic manipulations (like those computed in Sec.~\ref{Section II}) or by rearranging the asymptotic decomposition, but reflects the intrinsic UV structure of the integral.

The asymptotic regularization scheme makes this structure explicit by isolating the marginal sector from the rest of the integrand. 
In doing so, it separates the universal logarithmic divergence from the remaining contributions, which can be treated independently.
We now describe the regularization prescription that implements this separation in a systematic way.

We begin by adding and subtracting the asymptotic sectors inside the integrand of Eq.~\eqref{eq:I_radial_final},
\begin{align}
I = \Omega_D \int_0^\infty d\ell\, \ell^{D-1}
&\Big[
f(\ell)- f_{\mathrm{div}}(\ell)
+ f_{\mathrm{marg}}(\ell)+ f_{\mathrm{UV}}(\ell)
\Big].
\label{eq:split_identity}
\end{align}
This identity separates the integral into three asymptotically distinct contributions: a finite remainder coming from $f-f_{\mathrm{div}}$, a marginal sector, and a non-marginal UV sector.

Although the original integral is IR-finite, isolating the asymptotic sectors term by term may artificially generate spurious IR singularities. To control this intermediate decomposition, we introduce an auxiliary IR cutoff $\lambda>0$ and define
\begin{equation}
I=\lim_{\lambda\to0^+}
\left[
\widetilde I_{\mathrm{fin}}(\lambda)
+
\widetilde I_{\mathrm{marg}}(\lambda)
+
\widetilde I_{\mathrm{UV}}(\lambda)
\right],
\label{eq: principal I decomposition}
\end{equation}
where
\begin{align}
\widetilde I_{\mathrm{fin}}(\lambda)
&=
\Omega_D
\int_\lambda^\infty d\ell\,\ell^{D-1}\Big[f(\ell)- f_{\mathrm{div}}(\ell)
\Big],\label{eq:I_tilde_fin}\\
\widetilde I_{\mathrm{marg}}(\lambda)&=\Omega_D\int_\lambda^\infty d\ell\,\ell^{D-1}f_{\mathrm{marg}}(\ell), \label{eq:I_tilde_marg}
\\
\widetilde I_{\mathrm{UV}}(\lambda)&=\Omega_D\int_\lambda^\infty d\ell\,\ell^{D-1}f_{\mathrm{UV}}(\ell).\label{eq:I_tilde_uv}
\end{align}
Since $\lambda$ is introduced solely to render this decomposition well defined, the full result must be independent of it once the three sectors are recombined\footnote{Alternatively, we could have worked directly with \eqref{eq:split_identity} defining $f_\text{marg}(\ell)= b\,(\ell+\lambda)^{-D}$ in terms of a positive parameter $\lambda$ that renders this term IR-finite. The difference between this other approach and the one considered in the main text is only the shifting of a finite term from $\widetilde I_{\mathrm{fin}}(\lambda)$ to $\widetilde I_{\mathrm{marg}}(\lambda)$, as can be easily seen by introducing the change of variable $\ell\to\ell-\lambda$.}.

By construction, $\widetilde I_{\mathrm{fin}}(\lambda)$ is UV convergent. 
It contains the finite part of the original integral together with a spurious IR divergence as $\lambda \to 0^+$ which cancels with another IR divergence coming from $\widetilde I_{\mathrm{marg}}(\lambda)$ as we show in what follows. Therefore, after the removal of the IR cutoff, it contains all IR-sensitive and finite information of the original integral.

The structure of $\widetilde I_{\mathrm{UV}}(\lambda)$ depends on the precise asymptotic behavior of $f_{\mathrm{UV}}$. In complete generality, this sector may exhibit a variety of behaviors, potentially involving non-polynomial or non-local structures. In such cases, its treatment would require additional regularization techniques beyond the scope of the present work.
However, in all physically relevant situations known to us, including loop integrals in local QFT, effective field theories with local operators, and non-local theories with rational propagators, the UV sector is of algebraic (power-law) type. That is, there exist a finite integer $N$, coefficients $a_j$, and real exponents $p_0>p_1>\cdots>p_N>-D$, such that
\begin{equation}
f_{\mathrm{UV}}(\ell)=\sum_{j=0}^N a_j \ell^{p_j}.
\label{f_uv}
\end{equation}

In this case,
\begin{equation}
\widetilde I_{\mathrm{UV}}(\lambda)=\Omega_D\int_\lambda^\infty d\ell\,\ell^{D-1}\sum_{j=0}^N a_j \ell^{p_j}.
\end{equation}

Since the sum is finite, each monomial can be integrated term by term. Using the analytic continuation results established in Sec.~\ref{Section II}, each such contribution admits a unique continuation and vanishes in the limit $\lambda\to0^+$. Therefore,
\begin{equation}
\lim_{\lambda\to0^+}
\widetilde I_{\mathrm{UV, reg}}(\lambda)
=0.
\end{equation}

We conclude that, for every physically relevant case of algebraic UV behavior, the non-marginal sector does not contribute to the singular structure.

Assuming algebraic UV behavior as in \eqref{f_uv}, the regularized integral takes the form
\begin{align} \label{eq: I final}
I_{\mathrm{reg}}=\!\!\lim_{\lambda\to0^+}\Omega_D\!\int_\lambda^\infty \!\!\!d\ell\,\ell^{D-1}\big[f(\ell)-f_{\mathrm{div}}(\ell)\big]+\widetilde I_{\mathrm{marg,\, reg}}(\lambda),
\end{align}
where $\tilde I_{\mathrm{marg,\, reg}}(\lambda)$ denotes the regularized marginal integral.
Let us remind that the final result is independent of $\lambda$ and therefore we can safely take the limit $\lambda\to0^+$. 
Thus, asymptotic regularization isolates the only single pole, whose residue is precisely the coefficient $b$ of the marginal $\ell^{-D}$ term in the asymptotic expansion of the integrand. As discussed above, this structure encompasses all physically relevant UV behaviors known to us, namely those whose asymptotic expansion is of finite algebraic type. In such cases, which include the standard UV structures arising in local QFTs and effective field theories, the present method provides a complete and systematic identification of the physical divergence. 

\subsection{Regularization of the marginal integral}
\label{Section III.B}
 As can be read from \eqref{eq: f marg} and \eqref{eq:I_tilde_marg}, if present, the marginal integral takes the form
 \begin{equation} \label{I_marg}
    \widetilde I_\text{marg}(\lambda)
= \Omega_{D}\; b \int_\lambda^\infty d\ell\, \ell^{D-1} \ell^{-D} ,
\end{equation}
which, so far, we have not explicitly computed. Notice that this marginal sector cannot be simply regularized to zero like for the power-law contribution \eqref{eq:I_tilde_uv}, since it produces a pole (as shown in Sec.~\ref{Section II}). However, at this stage, the asymptotic regularization procedure has achieved its primary goal: it has isolated the only non-trivial UV divergent contribution of the integral, namely the marginal sector \eqref{I_marg}. As shown in Sec.~\ref{Section III.A}, this contribution is universal, being entirely determined by the coefficient $b$ of the $\ell^{-D}$ term in the asymptotic expansion of the integrand, and it captures the full logarithmic singular structure of the problem.

Importantly, up to this point no specific regulator has been used. The method itself does not rely on any particular prescription, but rather reduces the original problem to the evaluation of a single, structurally simple divergent integral. In this sense, asymptotic regularization separates the identification of the UV divergence from its explicit regularization.
If one wishes to obtain explicit, well-defined expressions, it is necessary to introduce a regulator to handle the marginal integral. This step is not unique, and different regularization schemes can be employed depending on the context and on the symmetries one aims to preserve. In fact, as shown in Appendix~\ref{app B}, the pole structure and the associated logarithmic dependence are universal and depend only on the marginal coefficient $b$, independently of the particular analytic regulator employed.

In the following, we illustrate this point by considering three standard choices: dimensional regularization, Pauli-Villars regularization, and Gaussian soft cutoff. 

\subsubsection{Dimensional regularization}
To implement the regularization, we analytically continue the dimension 
\begin{equation}
D \;\longrightarrow\; D-\varepsilon,
\qquad \Re\varepsilon>0,
\end{equation}
and take the limit $\varepsilon\to0$ at the end of the calculation.
Dimensional consistency requires the introduction of an arbitrary scale $\mu$
with the dimensions of $\ell$. Equation~\eqref{I_marg} then becomes 
\begin{equation}
\widetilde I_\text{marg}(\lambda)=\lim_{\varepsilon\to0} \widetilde J^\mathrm{dim}(\varepsilon,\lambda),
\end{equation}
with 
\begin{align}
  \widetilde J^\mathrm{dim}(\varepsilon,\lambda) &=
\mu^{\varepsilon}\,\Omega_{D-\varepsilon} 
\int_{\lambda}^\infty d\ell\, \ell^{D-1-\varepsilon} f_\mathrm{marg}(\ell) \nonumber\\
&=b \; \mu^{\varepsilon}\,\Omega_{D-\varepsilon}\int_{\lambda}^\infty d\ell\, \ell^{-1-\varepsilon}.
\label{eq:I_radial_final_reg}  
\end{align}

By continuing the dimension of the angular factor while keeping the asymptotic expansion of $f_\mathrm{marg}$ fixed, this divergence is regulated by $\varepsilon$. The volume of the unit $\Big(D-1-\varepsilon\Big)-$ sphere is
\begin{equation}
\Omega_{D-\varepsilon} = \frac{2\pi^{(D-\varepsilon)/2}}{\Gamma\!\left(\frac{D-\varepsilon}{2}\right)}.
\end{equation}
The marginal contribution \eqref{eq:I_radial_final_reg} can then be computed explicitly,
\begin{align}
  \widetilde J^\mathrm{dim \, reg}(\varepsilon,\lambda)&=
b\,\frac{2\pi^{(D-\varepsilon)/2}}{\Gamma\!\left(\frac{D-\varepsilon}{2}\right)}
\frac{1}{\varepsilon}
\Big(\frac{\mu}{\lambda}\Big)^\varepsilon \label{eq:general_pole_solved}  \\
&=b\,\Omega_D\Bigg[\frac{1}{\varepsilon}
+\log\frac{\mu}{\sqrt{\pi}\lambda} +\frac{1}{2}\psi^{(0)}\Big(\frac{D}{2}\Big)\Bigg]+\mathcal{O}(\varepsilon)\nonumber,
\end{align}
where $\psi^{(0)}$ denotes the digamma function \cite{PeskinSchroeder}. As anticipated, this marginal sector generates a genuine simple pole in $\varepsilon$, which encodes the intrinsic UV singularity of the integral. Also as mentioned above, it contains an IR divergent term proportional to $\log \lambda$ which cancels with the corresponding IR divergence in $\widetilde I_{\mathrm{fin}}(\lambda)$.

\subsubsection{Pauli--Villars regularization and scheme comparison}

In order to further illustrate the regulator-independent nature of the asymptotic regularization scheme, it is instructive to consider an alternative regularization of the marginal contribution based on Pauli-Villars (PV) subtraction.

Let us focus on the marginal integral given in \eqref{I_marg}, which can be written as
\begin{equation}
\widetilde I_\text{marg}(\lambda)=b\,\Omega_{D}\int_\lambda^\infty \frac{d\ell}{\ell}.
\end{equation}
This integral can be regularized by introducing a Pauli-Villars regulator $M$, leading to the modified expression
\begin{equation}
\widetilde I_\text{marg}(\lambda)=\lim_{M \to \infty }\widetilde J^\mathrm{PV}(M,\lambda),
\end{equation}
where
\begin{equation}
\widetilde J^\mathrm{PV}(M,\lambda)= b\,\Omega_{D}\int_\lambda^\infty d\ell \left(\frac{1}{\ell} - \frac{1}{\ell+M}\right).
\end{equation}
This integral can be computed explicitly, yielding
\begin{align}
  \widetilde J^\mathrm{PV}(M,\lambda)&= b\,\Omega_{D}
\log\frac{\lambda+M}{\lambda}\nonumber\\
&=b\,\Omega_{D}
\log\frac{M}{\lambda}+\mathcal{O}\Big(\frac{\lambda}{M} \Big),  
\end{align}
which makes explicit that the logarithmic divergence is now encoded in the regulator scale $M$, and emerges in the limit $M\to\infty$.

The Pauli-Villars implementation presented above is intended to illustrate the logarithmic structure of the marginal contribution and its relation to dimensional regularization. However, it has been applied at the level of the radial integral, after exploiting rotational invariance. In general, preserving gauge and Lorentz invariance requires introducing the regulator at the level of the full $D$-dimensional integrand, for instance through covariant Pauli-Villars subtraction acting on the propagators. The radial implementation discussed here should therefore be regarded as a simplified realization that captures the universal UV behavior, but does not by itself guarantee the preservation of gauge symmetries (unlike dimensional regularization).

Within the asymptotic regularization framework, this distinction is natural: the method isolates the universal marginal contribution independently of the regulator, while symmetry preservation is ensured by an appropriate choice of regularization scheme at the final stage.

This result might be compared with the expression obtained using dimensional regularization, where the marginal contribution produces a pole together with a logarithmic dependence on the renormalization scale $\mu$. In that case, the regularized result takes the form given in \eqref{eq:general_pole_solved}. The two schemes can be related by identifying the UV scales according to
\begin{equation}
\log M
\;\sim\;
\frac{1}{\varepsilon}
+
\log \mu,
\end{equation}
which corresponds to the formal relation $M = \mu\, e^{1/\varepsilon}$. This identification makes explicit that the pole in dimensional regularization encodes the same UV behavior as the logarithmic dependence on the Pauli-Villars mass.

Therefore, both regularization schemes lead to the same physical logarithmic divergence, with the same coefficient determined by the marginal term in the asymptotic expansion. The difference between them is entirely captured by scheme-dependent finite contributions.

\subsubsection{Gaussian soft cutoff}

As a further illustration of the regulator-independent nature of the asymptotic regularization scheme, it is useful to consider a smooth (soft) UV regulator. Instead of modifying the integrand by subtraction, one may introduce a multiplicative damping factor that suppresses the large-momentum region.

Focusing again on the marginal contribution \eqref{I_marg}, we consider the regularized integral
\begin{equation}
\widetilde I_\text{marg}(\lambda)
=\lim_{M\to\infty} \widetilde J^\mathrm{G}(M,\lambda),
\end{equation}
with
\begin{equation}
\widetilde J^\mathrm{G}(M,\lambda)
=
b\,\Omega_D
\int_\lambda^\infty \frac{d\ell}{\ell}\, e^{-\ell^2/M^2}.
\end{equation}
The gaussian factor provides a smooth suppression of the UV region controlled by the scale $M$.

This integral can be computed in closed form yielding
\begin{equation}
\widetilde J^\mathrm{G}(M,\lambda)
=\frac{b\,\Omega_D}{2}\,
\Gamma\!\left(0,\frac{\lambda^2}{M^2}\right),
\end{equation}
where $\Gamma\!\left(0,\lambda^2/M^2\right)$ denotes the incomplete gamma function \cite{Handbook}.

The UV behavior is obtained from the small-argument expansion. In the limit $M\to\infty$ one finds
\begin{equation}
\widetilde J^\mathrm{G}(M,\lambda)
=
b\,\Omega_D
\left[
\log\frac{M}{\lambda}
-\frac{\gamma_E}{2}\right]
+ \mathcal{O}\!\left(\frac{\lambda^2}{M^2}\right),
\end{equation}
with $\gamma_E$ the Euler-Mascheroni constant \cite{Handbook}. 

This expression makes explicit that the logarithmic divergence is again encoded in the UV scale $M$, exactly as in the Pauli-Villars case. The difference between the two schemes is entirely captured by finite terms, here involving the Euler-Mascheroni constant.

From the viewpoint of asymptotic regularization, this result is expected. The Gaussian factor acts as a multiplicative analytic regulator that modifies the integrand only in the UV, while leaving the asymptotic structure of the marginal term unchanged. As a consequence, the coefficient of the logarithmic divergence is still entirely determined by the marginal contribution $b\,\ell^{-D}$.

These previous examples illustrate that the role of asymptotic regularization is not to prescribe a specific regulator, but to isolate the universal UV structure of the integral. Once the marginal contribution has been identified, different regularization schemes can be implemented at this stage without affecting the physical content of the result.

\subsection{Control of non-local terms in a physical scale}
\label{Section III.C}

It is a standard result that, in the presence of a single physical scale, renormalized quantities in QFT generically develop non-local logarithmic terms on this scale \cite{Seeley1967,PeskinSchroeder}. Such contributions are typically derived using renormalization group arguments, dimensional regularization, or analyticity and unitarity properties of scattering amplitudes \cite{Cutkosky1960}. Here we recover this result from a different and more general viewpoint. The derivation does not rely on renormalization group methods, unitarity, or on the interpretation of the integral as a Feynman loop. Instead, the logarithmic term emerges solely from the asymptotic expansion of the integrand and the analytic regularization framework introduced above.

 As we will explicitly see in this subsection, whenever the asymptotic expansion contains a marginal term, a non-local logarithmic dependence on the physical scale is unavoidable, and its coefficient is entirely fixed by the UV behavior. The residue of the UV pole and the coefficient of the non-local logarithm therefore encode the same information, providing a direct and regulator-independent identification of the origin of logarithmic terms. This statement remains valid in situations where standard field-theoretic tools are unavailable, such as in QFT in curved spacetimes.

From a physical perspective, the logarithm in the physical scale arises when the UV subtraction is combined with integration over the IR region. While divergences originate in the UV, the non-local dependence on the physical scale  emerges through this interplay. Consequently, the same coefficient that multiplies the marginal scaling in the asymptotic expansion multiplies the logarithmic term after regularization.

To make these statements precise, consider an integral depending on a single scale $\Delta>0$, which for definiteness has dimensions of momentum $\ell$,
\begin{equation}
I(\Delta)=\Omega_D \int_0^\infty d\ell\, \ell^{D-1} f(\ell,\Delta).
\label{eq: Integral with scale}
\end{equation}
Assume that $f(\ell,\Delta)$ admits a large-momentum asymptotic expansion of the form \eqref{eq: f decomposition}, where the marginal contribution is $f_\mathrm{marg}(\ell,\Delta)=b(\Delta)\ell^{-D}$. 
We assume that there exists a dimensionless constant $\sigma_0>0$, independent of $\Delta$, such that the expansion is valid for $\ell \ge \sigma_0\Delta$, and that the remainder decays faster than $\ell^{-D}$ at large momentum.
Following the procedure described in Sec.~\ref{Section III.A}, we subtract the asymptotic contribution and consider the finite part given by \eqref{eq:I_tilde_fin}. To make the $\Delta$-dependence explicit, we consider the integral
\begin{align}
&I_{\mathrm{fin}}(\Delta)= \lim_{\lambda\to 0^+}\widetilde I_{\mathrm{fin}}(\lambda,\Delta)
 , \nonumber\\
 &\widetilde I_{\mathrm{fin}}(\lambda,\Delta)=
\Omega_D 
 \int_\lambda^{\infty}
d\ell\,\ell^{D-1}
\big[f(\ell,\Delta)-f_{\mathrm{div}}(\ell,\Delta)\big],
\label{eq:I finite scale}
\end{align}
and we split the finite part, $\widetilde I_{\mathrm{fin}}(\lambda,\Delta)$, at the scale $\sigma_0\Delta$ and analyze the two contributions separately:
\begin{align}
\widetilde I_{\mathrm{fin}}(\lambda,\Delta)&=\Omega_D 
\Bigg(
\int_\lambda^{\sigma_0\Delta}
d\ell\,\ell^{D-1}
\Big[f(\ell,\Delta)-f_{\mathrm{div}}(\ell,\Delta)\Big]
\nonumber\\
&+\int_{\sigma_0\Delta}^\infty
d\ell\,\ell^{D-1}
\Big[f(\ell,\Delta)-f_{\mathrm{div}}(\ell,\Delta)\Big]
\Bigg).
\label{eq: split integral}
\end{align}

In the first integral of \eqref{eq: split integral} all contributions are finite except the one coming from $f_{\mathrm{marg}}(\ell,\Delta)$, which diverges in the limit $\lambda\to0^+$. This contribution is
\begin{equation}
-b(\Delta)
\int_\lambda^{\sigma_0\Delta}
\frac{d\ell}{\ell}
=
-b(\Delta)\log\Big(\frac{\sigma_0\Delta}{\lambda}\Big).
\label{eq: marginal log Delta}
\end{equation}
The $\log\lambda$ term cancels against the UV $\log\lambda$ term generated by the regularization of the marginal sector (i.e.\ from $\widetilde I_\text{marg}(\lambda,\Delta)$).

On the other hand, in the second integral of \eqref{eq: split integral}, the asymptotic expansion \eqref{eq: f decomposition} is valid by construction. Hence,
\begin{align}
  &\int_{\sigma_0\Delta}^{\infty}
\!\!d\ell\,\ell^{D-1}
\big(f(\ell,\Delta)-f_{\mathrm{div}}(\ell,\Delta)\big)= 
\int_{\sigma_0\Delta}^{\infty}
\!\!d\ell\,\ell^{D-1}R(\ell,\Delta), 
\label{eq: int R}  
\end{align}
with $R(\ell,\Delta) <\mathcal{O}(\ell^{-D}),$ which is absolutely convergent and defines a smooth function  in $\Delta$. Since only an $\ell^{-1}$ behavior can generate a logarithmic primitive, the remainder cannot produce any additional $\log(\sigma_0\Delta)$ contribution that could cancel the logarithmic dependence generated in the first integral. Hence, whenever the asymptotic expansion of $f(\ell,\Delta)$ contains a marginal term $b(\Delta)\ell^{-D}$ with $b(\Delta)\neq0$, the integral \eqref{eq: split integral}  necessarily contains a contribution proportional to $\log(\Delta/\lambda)$, independently of the subleading details of the UV.

Furthermore, the previous analysis has been performed on the finite quantity $\widetilde I_\mathrm{fin}(\lambda,\Delta)$, which is UV-convergent. All divergences are encoded in $\widetilde I_\mathrm{marg}$  and $\widetilde I_\mathrm{UV}$. Consequently, $\widetilde I_\mathrm{fin}$  does not depend on the particular analytic regularization used to treat the UV sector. The logarithmic term is therefore independent of the regularization scheme employed.

It is important to emphasize that this analysis determines the functional form of the non-analytic dependence on the physical scale $\Delta$, but not its overall magnitude. However, once a single physical scale is present in the problem, physical considerations imply that this scale provides the natural argument of the logarithm \cite{PeskinSchroeder}. Any alternative choice differing by a dimensionless constant $\sigma_0$ would only modify the result by an additive constant, as we have seen, which corresponds to a local contribution that can be absorbed into local counterterms but the coefficient of $\log\Delta$ remains unchanged.

Since $\Delta$ is the only dimensionful parameter entering the problem, dimensional analysis implies that any contribution must be proportional to a power of $\Delta$ multiplied by a dimensionless function $h(\ell/\Delta)$,
\begin{equation}
f(\ell,\Delta)=\Delta^p\, h\!\left({\ell}/{\Delta}\right).
\end{equation}
This scaling behaviour follows directly from dimensional consistency and is consistent with the general structure of renormalized amplitudes in QFT, where the dependence on physical scales factorizes in the same way (see, e.g., \cite{Collins:105730}). The exponent $p$ controls the UV scaling of the integrand and is therefore related to the superficial degree of divergence of the integral~\eqref{eq: Integral with scale}. For $p<-D$ the integrand decays faster than $\ell^{-D}$ at large momentum and the integral becomes UV-convergent. Therefore, we will assume that $p\ge -D$. Dimensional consistency then requires that the same structure applies to $f_\mathrm{UV}$ and to the remainder $R$. For the marginal contribution $f_{\mathrm{marg}}=b(\Delta)\ell^{-D}$, dimensional consistency requires $b(\Delta)=b_0\,\Delta^{p+D}$ with $b_0\in\mathbb{R}$.

Applying this reasoning to the remainder $R(\ell,\Delta)$ in the second integral of \eqref{eq: split integral}, and performing the change of variables $y=\ell/\Delta$, the $\Delta$-dependence factorizes as $\Delta^{p+D}$, leaving a finite integral independent of $\Delta$. The same holds for contributions coming from $f(\ell,\Delta)$ and $f_\mathrm{UV}(\ell,\Delta)$, confirming that, when taking $\lambda \to 0$, they generate only  powers  of $\Delta$. The only exception is the marginal contribution $f_{\mathrm{marg}}$, whose primitive is logarithmic, as shown in \eqref{eq: marginal log Delta}.

Let us now consider the physically relevant situation in which $f(\ell,\Delta)$ is analytic in the physical scale $\Delta$. In this case, the dimension $p$ must be an integer and therefore $p+D$ must be a natural number. As a consequence, all contributions arising in \eqref{eq: Integral with scale}, except for the marginal one, reduce to integer powers of $\Delta$. Such terms correspond to local contributions and can be absorbed into local counterterms. The only non-–local dependence on $\Delta$ therefore originates from the marginal contribution displayed in \eqref{eq: marginal log Delta}. 

An important consequence of this analysis is that the coefficient of the logarithmic term is uniquely fixed by the UV asymptotics of the integrand. In particular, both the UV divergence and the logarithmic dependence on the physical scale originate from the same marginal contribution $f_{\mathrm{marg}}(\ell,\Delta)=b(\Delta)\ell^{-D}$, with $b(\Delta)=\Delta^{p+D} b_0$. As a result, the coefficient multiplying the UV pole is precisely the same coefficient that multiplies the logarithmic term in $\Delta$. 

\subsection{An alternative derivation of the non-local logarithms}
\label{Section III.D}

In Sec.~\ref{Section III.C} we showed that, whenever the marginal contribution $f_{\mathrm{marg}}$ is present, the resulting dependence on the physical scale $\Delta$ takes the form
\begin{equation}
    I_{\mathrm{fin}}(\Delta) \sim \Delta^{p+D}\,b_0\,\log\Delta .
\end{equation}
We now derive the same result from a complementary viewpoint that makes the control of the scale dependence particularly transparent.

Let us consider the finite integral $\widetilde I_{\mathrm{fin}}(\lambda,\Delta)$ defined in~\eqref{eq:I finite scale}. When $\Delta$ is the only physical scale in the problem, dimensional analysis implies that the functions appearing in the integrand can be written as
\begin{equation}
   f(\ell,\Delta)=\Delta^p\,h(\ell/\Delta), 
\quad 
f_{\mathrm{UV}}(\ell,\Delta)=\Delta^p\,h_{\mathrm{UV}}(\ell/\Delta), 
\end{equation}
while the marginal term takes the form
\begin{equation}
   f_{\mathrm{marg}}(\ell,\Delta)=\Delta^{p+D}\,b_0\,\ell^{-D}, 
\end{equation}
where $h$ and $h_{\mathrm{UV}}$ are dimensionless functions. As discussed in the previous subsection, these expressions simply reflect the fact that $\Delta$ is the only dimensionful parameter entering the integrand.

Introducing the dimensionless variable $y=\ell/\Delta$, the finite contribution becomes
\begin{align}
      \widetilde I_{\mathrm{fin}}(\lambda,\Delta)
   &= \Delta^{p+D}\,\Omega_D 
   \int_{\lambda/\Delta}^{\infty}
   dy\,y^{D-1}
   \Big[
   h(y) \nonumber\\
   &\quad -h_{\mathrm{UV}}(y)-b_0\,y^{-D}
   \Big]. 
\end{align}

The explicit dependence on $\Delta$ can be extracted by differentiating with respect to the scale. A straightforward computation gives
\begin{align}
      \Delta \frac{d\widetilde I_{\mathrm{fin}}(\lambda,\Delta)}{d\Delta}
    &=
    (p+D)\,\widetilde I_{\mathrm{fin}}(\lambda,\Delta)
     \nonumber\\
    &\quad-\Delta^{p+D}\Omega_D\,b_0
    \Big[
    1+\mathcal{O}\!\left(\frac{\lambda}{\Delta}\right)
    \Big].
    \label{appb: differential equation}  
\end{align}
 This relation has the structure of a renormalization-group equation for the scale $\Delta$, where the marginal coefficient $b_0$ acts as the source responsible for the logarithmic running.

Equation \eqref{appb: differential equation} provides a first-order differential equation governing the scale dependence of $\widetilde I_{\mathrm{fin}}(\lambda,\Delta)$. To determine the solution one must specify an initial condition. Dimensional consistency restricts the possible choices to scales proportional to the IR cutoff $\lambda$. We therefore fix the value of the integral at $\Delta=\lambda/\sigma$, where $\sigma$ is a dimensionless constant. Solving \eqref{appb: differential equation} with this condition and taking the limit $\lambda\to0^+$, while retaining the only divergent contribution, yields
\begin{equation}
     I_{\mathrm{fin}}(\Delta)
     =
     -\Delta^{p+D}\,\Omega_D\,b_0
     \log\frac{\Delta}{\sigma_0\lambda},
     \qquad
     \sigma_0\in\mathbb{R}.
     \label{log of the escale}
\end{equation}

This expression coincides with the result obtained in Sec.~\ref{Section III.B}, confirming that the logarithmic dependence on the physical scale $\Delta$ is entirely dictated by the marginal contribution in the asymptotic expansion.

\subsection{Example}
\label{Section III.E}

Let us illustrate the application of the asymptotic regularization scheme with a standard class of integrals appearing in QFT.
We consider the following integrals, which arise in the one-loop radiative correction to the photon propagator in QED \cite{PeskinSchroeder}:
\begin{equation}
    I=\int_{\mathbb{R}^4} \frac{d^4\ell\,\ell^{2a}}{ (2\pi)^4(\ell^2+\Delta^2)^2},
\end{equation}
where $a$ takes the values $0$ and $1$.

The large-$\ell$ asymptotic expansion of the integrand reads
\begin{equation}
    \frac{\ell^{2a}}{ (\ell^2+\Delta^2)^2} \sim \frac{a}{\ell^2}+(-1)^a\frac{(a+1)\Delta^{2a}}{\ell^4}+\mathcal{O}\Big(\frac{\Delta^{2(a+1)}}{\ell^{6}}\Big).
\end{equation}
We now apply the asymptotic regularization prescription. To make the calculation explicit, we use dimensional regularization to treat the marginal sector as shown in Sec.~\ref{Section III.B}. Then, performing the dimensional continuation, introducing an auxiliary infrared cutoff $\lambda$, separating the integral according to the asymptotic decomposition, combining the finite contribution with the regularized logarithmic term, and taking the limit $\lambda\to0^+$, the regularized integral reads
\begin{equation}
    I_{\mathrm{reg}}=\frac{(a+1)\Delta^{2a}}{(4\pi)^2} \Big[(-1)^a\frac{2}{\varepsilon}-\log\frac{4\pi\mu^2}{\Delta^2}-\frac{a}{2}-\gamma_E \Big]+\mathcal{O}(\varepsilon).
\end{equation}

Notice that some constant factors differ from the general expressions in \eqref{eq:general_pole_solved}. This originates from the fact that in the present example we have analytically continued the dimension also in the constant factor of the denominator. Such differences can be absorbed into a redefinition of the renormalization scale $\mu$.

For comparison, the same integral evaluated using standard dimensional regularization gives the same result  \cite{PeskinSchroeder}.

\section{Comparison with others regularization schemes}
\label{Section IV}

Up to this point we have introduced the asymptotic regularization scheme, established its general structure, and illustrated its application through explicit examples. Since the method employs other regulators to treat the marginal sector, one question arises: how does it differ from the standard regularization schemes?

The purpose of this section is to clarify this relation. We first show that, when implemented through dimensional continuation, the asymptotic scheme preserves the same symmetry properties that make dimensional regularization particularly powerful, such as gauge invariance and Lorentz covariance. We then emphasize a conceptual advantage of the asymptotic approach: the pole structure of the integral can be identified directly from the UV scaling of the integrand, without any particular regulator. Finally, we present an example where other regularization schemes cannot be applied in its standard form, while the asymptotic subtraction procedure remains well defined.

\subsection{Symmetries preservation}
\label{Section IV.A}

 The asymptotic regularization scheme introduced in this work may employ analytic continuation of the spacetime dimension as a regulator for the marginal integral. Applied in that way, it shares a common foundation with dimensional regularization. The continuation acts at the level of the angular factor, while the UV singular sector is isolated through the large-momentum asymptotic expansion of the integrand. Although, as discussed in Sec.~\ref{Section III.B}, the marginal term could in principle be treated using an arbitrary regulator, the choice of the spacetime dimension $D$ as regulator has some advantages. Indeed, implementing the subtraction through dimensional continuation ensures that gauge symmetry and Lorentz invariance are preserved, aligning the procedure with the structural advantages of dimensional regularization. Certainly, Ward and Slavnov--Taylor identities remain valid, as no hard cutoff or non-covariant scale is introduced. Lorentz invariance is likewise manifest, since the procedure depends only on rotationally invariant combinations such as $\ell^2$ and does not single out preferred directions. In covariant formulations, the method is compatible with diffeomorphism invariance.

We recall that the essential difference between these two schemes lies in the way the UV sector is identified.

\subsection{ Direct identification of the UV pole}
\label{Section IV.B}

 Within the asymptotic framework, the presence or absence of UV poles can be determined directly from the large-momentum expansion of the integrand. In particular, logarithmic divergences arise only when a marginal contribution is present in the asymptotic expansion.
To illustrate this point we consider an integral arising in models with MDRs of Corley-Jacobson superluminal type \cite{corley1996, LopezNacir:2009bhs},
\begin{align}
  I&=\int_{\mathbb{R}^3} \frac{d^3 \ell}{\sqrt{{\ell^4}/{\kappa_c^2}+\ell^2+m^2}} \nonumber\\
&= \Omega_3\int_0^\infty d\ell\,\frac{\ell^2}{\sqrt{{\ell^4}/{\kappa_c^2}+\ell^2+m^2}},  
\end{align}
where $\kappa_c$ sets the UV scale.

The large-$\ell$ expansion of the integrand is
\begin{equation}
\frac{1}{\sqrt{{\ell^4}/{\kappa_c^2}+\ell^2+m^2}}
= \frac{\kappa_c}{\ell^2} \bigg[1+ \mathcal{O}\!\left(\frac{\kappa_c^2}{\ell^{2}}\right)\bigg],
\qquad (\ell\to\infty).
\end{equation}
Importantly, no marginal contribution of order $\ell^{-3}$ appears. According to the general structure established in Sec.~\ref{Section II}, this immediately implies the absence of a logarithmic divergence.
The regularized expression is obtained by subtracting the leading asymptotic term,
\begin{equation}
I_{\mathrm{reg}}=\Omega_3\int_0^\infty d\ell
\left(
\frac{\ell^2}{\sqrt{{\ell^4}/{\kappa_c^2}+\ell^2+m^2}}
- \kappa_c
\right),
\end{equation}
which is finite in three spatial dimensions without the need of introducing any regulator.

Notice that other regularization schemes can, in principle, be applied to this integral. However, within that framework the absence of poles becomes manifest only after performing the specific regularization. In contrast, within the asymptotic approach the conclusion follows directly from the UV expansion itself. This fact becomes particularly important in situations where other regularization schemes cannot be implemented in a straightforward way using the well-known expressions. 
In such cases the asymptotic subtraction procedure can be useful, as we illustrate in the following example.

\subsection{Extension beyond standard regularization schemes}
\label{Section IV.C}

Our asymptotic regularization can be applied in situations where standard regulator-based approaches are not naturally adapted. As an illustrative example, consider the momentum integral associated with the Unruh-type dispersion relation \cite{Unruh1976db} in $D$ spatial dimensions
\begin{equation}
I(D) = \Omega_D \int_0^\infty 
\frac{\ell^{D-1}\, d\ell}{\sqrt{\kappa_c^2 \tanh^2(\ell/\kappa_c) + m^2}},
\label{eq:UnruhIntegral}
\end{equation}
where $\kappa_c$ is a fixed UV scale.

At the level of the full integral, the implementation of standard regularization schemes is not straightforward. In particular, regulator-based prescriptions such as Pauli--Villars or smooth cutoffs can be formally introduced, but they act globally on the integrand and do not naturally separate the UV structure from the IR behavior. As a consequence, they obscure the identification of the genuinely singular contributions and introduce regulator-dependent modifications over the entire integration domain.

A more severe obstruction arises in dimensional regularization. In the UV limit $\ell \to \infty$, the integrand behaves as $\ell^{D-1}$, which requires $\mathrm{Re}(D)<0$ for convergence. In the IR limit $\ell \to 0$, the same scaling implies convergence only if $\mathrm{Re}(D)>0$. Therefore, there is no open region in the complex $D$-plane where the integral converges absolutely. Since dimensional regularization relies on the existence of such a domain to define the analytic continuation of the full integral, it cannot be directly implemented for integrals of the form \eqref{eq:UnruhIntegral}.

One may attempt to circumvent this obstruction by performing asymptotic expansions and applying dimensional regularization to the resulting contributions. However, such procedures do not correspond to a direct implementation of dimensional regularization on the original integral, but rather to the introduction of intermediate decompositions, typically involving additional scales or cutoffs.

The asymptotic regularization method bypasses these difficulties. Its implementation does not require introducing a regulator at the level of the full integral, nor does it rely on the existence of a convergence domain in complex dimension. Instead, the method applies a specific regulator only for the isolated marginal sector.

In the present case, the large-momentum expansion, when $\ell\to\infty$, reads 
\begin{align}
   \frac{1}{\sqrt{\kappa_c^2 \tanh^2(\ell/\kappa_c) + m^2}} = \frac{1}{\sqrt{\kappa_c^2 + m^2}}\big[1
+ \mathcal{O}(e^{-2\ell/\kappa_c})\big]. 
\end{align}
The leading UV contribution is a constant. Subtracting this term,
\begin{align}
    I_{\mathrm{reg}}=\Omega_D\int_0^\infty \!\!\!d\ell \, \ell^{D-1}
\Bigg[ &\frac{1}{\sqrt{\kappa_c^2 \tanh^2(\ell/\kappa_c) + m^2}}
\nonumber\\
&-\frac{1}{\sqrt{\kappa_c^2 + m^2}}
\Bigg],  
\end{align}
produces an integrand that decays exponentially at large momentum and remains regular in the IR. The resulting integral is finite directly in $D$ dimensions.

In particular, no marginal term appears in the asymptotic expansion, and therefore no logarithmic divergence or pole structure is generated. 

This illustrates a key advantage of the asymptotic regularization scheme: divergences are identified and isolated at the level of the asymptotic structure, rather than controlled through a global modification of the integrand.

\section{Conclusions}
\label{Section V}

In this work we have presented a constructive formulation of a regularization scheme for UV divergences based on the large-momentum expansion of the integrand. The central idea is that the singular UV sector can be identified directly from the asymptotic structure of the integrand, prior to introducing any regulator. In this framework, the integrand naturally separates into three distinct contributions associated with the asymptotic expansion: power-divergent contributions, a marginal contribution, and a rapidly decaying remainder. While the procedure is completely general and can be applied to arbitrary asymptotic structures, in this work we have focused on the physically most relevant situation in QFT, namely integrals exhibiting algebraic (power-law) UV behavior.

For this class of integrals we have shown that the UV singular structure is remarkably simple. The power-divergent contributions correspond to local terms that can be absorbed into counter-terms and therefore, can be set equal to zero using analytical continuation, whereas the rapidly decaying remainder generates finite contributions. The only genuine UV singularity arises from the marginal term in the asymptotic expansion. In this sense, the UV singular structure of the integral can be read directly from the asymptotic expansion of the integrand.

A direct consequence of this structure is the universal appearance of a logarithmic dependence on a physical scale when it is the only scale present in the analysis. The logarithmic term originates from the primitive of the marginal contribution and reflects the interplay between the UV subtraction and the IR scale entering the problem. In particular, the coefficient governing the logarithmic running coincides with the coefficient  of the isolated UV divergence. This relation follows directly from the asymptotic structure of the integrand: both the pole and the logarithmic dependence originate from the same marginal contribution in the large-momentum expansion, in close analogy with the structure underlying renormalization-group running.

From a broader perspective, the present framework is related in spirit to subtraction schemes that exploit the short-distance structure of QFT, such as BPHZ renormalization~\cite{Bogoliubov:1957gp}, adiabatic subtraction in curved spacetimes \cite{parker_toms_qft_curved_spacetime}, or the Hadamard prescription \cite{Ward:1950xp}. In those approaches the divergent contributions are removed through local counterterms determined by the short-distance behavior of the theory. The asymptotic scheme developed here follows a different organizing principle: the UV sector is first decomposed according to the large-momentum expansion of the integrand, which isolates the marginal contribution as the unique source of logarithmic running. In this way, the pole structure and the associated logarithmic dependence are identified directly from the UV scaling properties of the integrand.

The asymptotic regularization scheme provides some relatable advantages. When the dimension is employed as regulator for the marginal sector, the present scheme inherits the main structural advantages of dimensional regularization. Gauge symmetry, Lorentz invariance~, and, in covariant settings, diffeomorphism invariance are preserved since the procedure does not introduce hard cutoffs or non-covariant scales. In our asymptotic scheme the UV singular sector is identified directly from the large-momentum behavior of the integrand, so that the pole structure can be read off immediately from the existence of a marginal term.
More importantly, the method can be applied even in situations where other regularization schemes cannot be implemented in their standard form. Since the action of a concrete regulator within  asymptotic regularization scheme is only required for the marginal sector, it remains applicable in such cases. This suggests that the asymptotic approach provides a natural framework to analyze UV singularities in situations where standard regularization methods are difficult to implement.

The results presented here suggest that asymptotic regularization should be viewed not as a reformulation of any well-known regularization scheme, but as a subtraction scheme determined directly by the asymptotic behavior of the integrand. Its applicability is therefore controlled by the UV scaling of the theory rather than the choice of a particular regulator.

Future work will explore the application of this framework to quantum field theory in curved spacetimes in presence of MDRs. In such settings the UV structure of momentum integrals often departs from the standard relativistic form, and the asymptotic approach developed here may provide a particularly natural tool to identify and isolate the relevant divergences.

\acknowledgments

We are very grateful to Juan José Sanz Cillero for his valuable guidance and insightful discussions on the interpretation, formal aspects, and applicability of the regularization method presented in this work.
This work was financially supported by Grant PID2023-149018NB-C44 funded by
MCIN/AEI/10.13039/501100011033 and by ``ERDF/EU''.
RN was funded by the Royal Society through the University Research Fellowship Renewal URF$\backslash$R$\backslash$221005. CDR acknowledges financial support from
MCIU (Ministerio de Ciencia, Innovación y Universidades, Spain) fellowship FPU24/03217.

\appendix

\section{Asymptotic regularization for IR-divergent integrals}
\label{app A}

The asymptotic regularization procedure described in Sec.~\ref{Section III.A} can be extended straightforwardly to the treatment of IR singularities. The logic is entirely analogous, with the only difference that the relevant asymptotic region is now $\ell \to 0$ rather than $\ell \to \infty$.

In fact, this correspondence can be made explicit by the change of variables $u=1/\ell$, which formally maps the IR region $\ell\to0$ into a UV regime $u\to\infty$. Although we will not rely on this transformation in what follows, it provides an intuitive way to understand the parallelism between the two cases.

Let us consider again a rotationally invariant $D$-dimensional integral of the form~\eqref{eq:I_radial_final} and assume that the integrand is UV finite. The potential singular behavior is therefore localized in the IR region. Independently of the detailed structure of $f$, its small-$\ell$ behavior admits an asymptotic decomposition of the form
\begin{equation}
    f(\ell) \sim f_{\text{div}}^{\text{IR}}(\ell) + R(\ell),
\end{equation}
where $f_{\text{div}}^{\text{IR}}$ collects all IR-divergent contributions, and $R(\ell)$ is regular as $\ell \to 0$. As in the UV case, we further decompose
\begin{equation}
    f_{\text{div}}^{\text{IR}}(\ell)
    =
    f_{\mathrm{IR}}(\ell)
    +
    f_{\mathrm{marg}}(\ell),
\end{equation}
where $f_{\mathrm{IR}}$ contains non-marginal contributions and, if present, the marginal term takes the universal form given by \eqref{eq: f marg}.

As in the UV analysis, the logarithmic singularity is uniquely associated with the marginal scaling $\ell^{-D}$, while all other contributions lead either to finite terms or to divergences of a different (non-logarithmic) nature.

We now implement the asymptotic subtraction at the level of the integral. Adding and subtracting the divergent sectors. In this case, isolating the asymptotic contributions may generate spurious UV divergences in the intermediate expressions. To control this, we introduce an auxiliary UV cutoff $\Lambda$ and define
\begin{equation}
I=\lim_{\Lambda\to\infty}
\left[
\widetilde I_{\mathrm{fin}}(\Lambda)
+
\widetilde I_{\mathrm{marg}}(\Lambda)
+
\widetilde I_{\mathrm{IR}}(\Lambda)
\right],
\end{equation}
with
\begin{align}
\widetilde I_{\mathrm{fin}}(\Lambda)
&=
\Omega_D
\int_0^\Lambda d\ell\,\ell^{D-1}
\Big[f(\ell)- f_{\text{div}}^{\text{IR}}(\ell)\Big],\\
\widetilde I_{\mathrm{marg}}(\Lambda)
&=
\Omega_D
\int_0^\Lambda d\ell\,\ell^{D-1} f_{\mathrm{marg}}(\ell),\\
\widetilde I_{\mathrm{IR}}(\Lambda)
&=
\Omega_D
\int_0^\Lambda d\ell\,\ell^{D-1} f_{\mathrm{IR}}(\ell).
\end{align}

By construction, $\widetilde I_{\mathrm{fin}}(\Lambda)$ is IR convergent. It contains the finite contribution of the original integral together with a spurious UV divergence as $\Lambda \to \infty$, which cancels against a corresponding contribution in $\widetilde I_{\mathrm{marg}}(\Lambda)$.

The structure of $\widetilde I_{\mathrm{IR}}(\Lambda)$ depends on the detailed small-$\ell$ behavior of $f_{\mathrm{IR}}$. In all physically relevant situations, this sector is of algebraic (power-like) type, i.e.
\begin{equation}
f_{\mathrm{IR}}(\ell)=\sum_{j=0}^N a_j \ell^{p_j},
\qquad p_j < -D.
\end{equation}
In this case, each contribution can be treated using the analytic continuation methods developed in Sec.~\ref{Section II}. Although those results were originally derived for positive powers, they extend to the present situation \emph{mutatis mutandis} by analytic continuation. As a consequence,
\begin{equation}
\lim_{\Lambda\to\infty}
\widetilde I_{\mathrm{IR, reg}}(\Lambda)
=0.
\end{equation}
Therefore, as in the UV case, the non-marginal sector does not contribute to the singular structure.

The only non-trivial contribution arises from the marginal term, which takes the form
\begin{equation}
\widetilde I_{\mathrm{marg}}(\Lambda)
=
b\,\Omega_D
\int_0^\Lambda \frac{d\ell}{\ell}.
\end{equation}
This integral encodes the full logarithmic IR divergence of the original expression.

At this stage, the asymptotic regularization procedure has again reduced the problem to the evaluation of a single universal divergent contribution. Its regularization can be carried out using any convenient scheme, such as dimensional regularization or Pauli--Villars, exactly as discussed in Sec.~\ref{Section III.B}. The resulting logarithmic structure and its coefficient are independent of this choice.

In summary, the asymptotic regularization framework applies equally to IR singularities: the divergent structure is entirely captured by the marginal $\ell^{-D}$ term in the small-momentum expansion of the integrand, while all other contributions can be treated unambiguously and do not affect the logarithmic behavior.

As a final remark, it is worth emphasizing that the marginal contribution 
$f_{\mathrm{marg}}$ has exactly the same structure in the UV and IR analyses. 
As a consequence, the results obtained in Sec.~\ref{Section III.D} apply to the IR case discussed here.  
In particular, whenever the integrand depends on a single physical scale $\Delta$, the presence of a marginal term necessarily induces a non-local logarithmic dependence of the form \eqref{log of the escale} with a coefficient  $b_0$ entirely fixed by the same marginal term that controls the divergence. 
Therefore, independently of whether the singular behavior originates in the UV or in the IR region, both the divergence and the logarithmic dependence on the physical scale are universally determined by $f_{\mathrm{marg}}$. The difference between the two cases is purely kinematical, while the underlying structure remains identical.

\section{Analytic scheme independence and control of the running}
\label{app B}

We now clarify in which sense the UV singular structure identified through asymptotic regularization is independent of the particular analytic regulator employed, and why the coefficient of the marginal asymptotic term is the only quantity that controls the physical running.

In Sec.~\ref{Section III} we saw that, for non-gauge theories, Pauli--Villars regularization and dimensional regularization are equivalent. Here we consider a broad class of analytic multiplicative regulators. Let $G(\ell;\varepsilon)$ be a one-parameter family, analytic in $\varepsilon$ in a neighborhood of $\varepsilon=0$, such that in the UV
\begin{equation}
G(\ell;\varepsilon)
=
\mu^{\varepsilon}\,\ell^{-\varepsilon}
A(\varepsilon)\,\bigl(1+r_\varepsilon(\ell)\bigr),
\qquad \ell\to\infty,
\label{eq:G_uv}
\end{equation}
where $\mu$ is a renormalization scale, $A(\varepsilon)$ is analytic near $\varepsilon=0$ with $A(0)=1$, and $r_\varepsilon(\ell)\to0$ as $\ell\to\infty$. We assume that for sufficiently large $\Re\varepsilon$ the regulated integral is convergent and admits an analytical continuation to $\varepsilon=0$.

Consider a function $f(\ell)$ admitting an asymptotic expansion of algebraic type, where $f_\mathrm{UV}$ is given by \eqref{f_uv}, and let $b$ denote the coefficient of the marginal term $f_{\mathrm{marg}}(\ell)=b\,\ell^{-D}$. The marginal integral \eqref{I_marg} can be regularized using any analytic regulator of the family $G(\ell;\varepsilon)$, leading to
\begin{equation}
    \widetilde I_\mathrm{marg}(\lambda)=\lim_{\varepsilon\to 0} J^A_\mathrm{reg}(\varepsilon,\lambda),
\end{equation}
where
\begin{equation}
\widetilde J^A_\mathrm{reg}(\varepsilon,\lambda)
=
b\,\Omega_D
\int_\lambda^\infty d\ell\,
\ell^{D-1}\,
G(\ell;\varepsilon),
\end{equation}
where the only possible source of a singularity occurs at $\varepsilon=0$.
Using \eqref{eq:G_uv}, we obtain
\begin{equation}
\widetilde J^A_\mathrm{reg}(\varepsilon,\lambda)
=b\,\mu^{\varepsilon} A(\varepsilon)
\int_\lambda^\infty
d\ell\, \ell^{-1-\varepsilon}
\Big[1+r_\varepsilon(\ell)\Big].
\end{equation}
Expanding near $\varepsilon=0$, one finds
\begin{equation}
\widetilde J^A_\mathrm{reg}(\varepsilon,\lambda)
=A(0)\,b\,\Omega_D\Bigg[\frac{1}{\varepsilon}
+\log\frac{\mu}{\sqrt{\pi}\lambda} +\frac{1}{2}\psi^{(0)}\Big(\frac{D}{2}\Big)\Bigg]+\mathcal{O}(\varepsilon),
\label{eq:pole structure}
\end{equation}
which coincide with \eqref{eq:general_pole_solved}. Notice that in the previous expansion, we have used that the contribution involving $r_\varepsilon(\ell)$ is analytic at $\varepsilon=0$ and that $A(0)=1.$

The simple pole and the associated logarithmic dependence on the renormalization scale $\mu$ are therefore entirely determined by the marginal coefficient $b$, as discussed in Sec.~\ref{Section III}.

Consequently, within MS type schemes \cite{Collins:105730, PeskinSchroeder}, the running with respect to $\mu$ is fixed entirely by the marginal asymptotic coefficient $b$, independently of the specific analytic realization of the regulator. The UV structure relevant for renormalization group flow is therefore universal and controlled uniquely by the $\ell^{-D}$ term in the asymptotic expansion.

\bibliography{biblio}

\end{document}